\documentclass[onecolumn,amsmath,showpacs,nofootinbib,11pt]{revtex4-2}
\usepackage{graphicx}
\usepackage{dcolumn}
\usepackage{bm}
\usepackage{color} 
\usepackage{slashed}
\begin{document}
\newcommand{\hs}{\hspace*{0.5cm}}
\newcommand{\vs}{\vspace*{0.5cm}}
\newcommand{\be}{\begin{equation}}
\newcommand{\ee}{\end{equation}}
\newcommand{\bea}{\begin{eqnarray}}
\newcommand{\eea}{\end{eqnarray}}
\newcommand{\ben}{\begin{enumerate}}
\newcommand{\een}{\end{enumerate}}
\newcommand{\bde}{\begin{widetext}}
\newcommand{\ede}{\end{widetext}}
\newcommand{\nn}{\nonumber}
\newcommand{\crn}{\nonumber \\}
\newcommand{\Tr}{\mathrm{Tr}}
\newcommand{\non}{\nonumber}
\newcommand{\noi}{\noindent}
\newcommand{\al}{\alpha}
\newcommand{\la}{\lambda}
\newcommand{\bet}{\beta}
\newcommand{\ga}{\gamma}
\newcommand{\va}{\varphi}
\newcommand{\om}{\omega}
\newcommand{\pa}{\partial}
\newcommand{\+}{\dagger}
\newcommand{\fr}{\frac}
\newcommand{\bc}{\begin{center}}
\newcommand{\ec}{\end{center}}
\newcommand{\Ga}{\Gamma}
\newcommand{\de}{\delta}
\newcommand{\De}{\Delta}
\newcommand{\ep}{\epsilon}
\newcommand{\varep}{\varepsilon}
\newcommand{\ka}{\kappa}
\newcommand{\La}{\Lambda}
\newcommand{\si}{\sigma}
\newcommand{\Si}{\Sigma}
\newcommand{\ta}{\tau}
\newcommand{\up}{\upsilon}
\newcommand{\Up}{\Upsilon}
\newcommand{\ze}{\zeta}
\newcommand{\ps}{\psi}
\newcommand{\Ps}{\Psi}
\newcommand{\ph}{\phi}
\newcommand{\vph}{\varphi}
\newcommand{\Ph}{\Phi}
\newcommand{\Om}{\Omega}
\newcommand{\AdrHEPC}{Phenikaa Institute for Advanced Study and Faculty of Basic Science, Phenikaa University, Yen Nghia, Ha Dong, Hanoi 100000, Vietnam}
\newcommand{\AdrH}{Institute of Physics, Vietnam Academy of Science and Technology, 10 Dao Tan, Ba Dinh, Hanoi 100000, Vietnam}

\title{Can the Higgs field feel a dark force?} 

\author{Phung Van Dong} 
\email{Corresponding author; dong.phungvan@phenikaa-uni.edu.vn}
\affiliation{\AdrHEPC} 
\author{Duong Van Loi}
\email{loi.duongvan@phenikaa-uni.edu.vn}
\affiliation{\AdrHEPC}
\author{Do Thi Huong}
\email{dthuong@iop.vast.vn}
\affiliation{\AdrH}

\date{\today}

\begin{abstract}

We argue that if an electroweak Higgs field possesses a dark gauge charge responsible for dark matter stability, the $W$-boson mass deviation is properly induced, besides appropriately generated neutrino masses. We examine a simple model in which the usual Higgs doublet plays the role but dark matter candidates are somewhat input by ad hoc. We look for a realistic model that fully realizes such observation, thereby neutrino mass and dark matter are naturally supplied by a dark non-abelian gauge symmetry.   
                  
\end{abstract} 

\maketitle

\section{Motivation} 

Neutrino mass \cite{Kajita:2016cak,McDonald:2016ixn} and dark matter \cite{Jungman:1995df,Bertone:2004pz,Arcadi:2017kky} are two among numerous open-questions in science which require new physics beyond the standard model. The $W$-boson mass anomaly recently measured by CDF collaboration with very high precision at 7$\sigma$ \cite{CDF:2022hxs} might reveal new insight into the physics of neutrino mass and dark matter. This work is aimed at finding this correlation.     

Stability of dark matter may be manifestly preserved by a dark gauge symmetry under which dark matter is charged. If a Higgs doublet $H$ also carries a dark charge, normalized to 1 for brevity, it couples to dark gauge boson, called $X$, as $\mathcal{L}\supset  -ig_{X} H^\dagger X^\mu D_\mu H+H.c.$, where $g_{X}$ is the dark gauge coupling, and $D_\mu$ is the standard model covariant derivative. Assuming $X$ above the weak scale $v_{\mathrm{w}}\equiv 246$ GeV, integrating it out leads to an effective interaction, \be \mathcal{L}_{\mathrm{eff}}\supset -\fr{2g^2_{X}}{m^2_{X}}|H^\dagger D_\mu H|^2,\label{eq}\ee where $m_{X}$ is the $X$ mass concerning dark charge breaking, and 2 comes from $s,t$-channel contribution. This effective interaction contributes to the $T$-parameter at tree level as $T_{\mathrm{tree}}=v^4g^2_{X}/\al v^2_{\mathrm{w}} m^2_{X}$, where $v=\sqrt{2}\langle H^0\rangle$ and $\al=1/128$. The $W$-boson mass anomaly is properly solved, given that $T\simeq 0.145$ \cite{Strumia:2022qkt}, acquiring $m_{X}/g_{X} \simeq 5(v/v_{\mathrm{w}})^2$ TeV. 

If $H$ belongs to a new physics, it implies an extra 1-loop contribution to $T$ through its coupling to electroweak gauge bosons $\mathcal{L}\supset (D^\mu H)^\dagger (D_\mu H)$, such as  
\be T_{\mathrm{rad}} = \fr{f(m^2_1,m^2_2)}{8\pi^2 \al v^2_{\mathrm{w}}}\simeq \fr{(\Delta m)^2}{4\pi^2 \al v^2_{\mathrm{w}}},\ee where $f(x,y)=(x+y)/2-xy\ln(x/y)/(x-y)$ and the $H$ mass splitting $\Delta m=m_{1}-m_2\ll m_{1,2}$. This result agrees with the $W$-boson mass anomaly if $\Delta m\simeq 52$ GeV. This loop effect is especially significant if $v=0$, i.e. $T_{\mathrm{tree}}=0$, due to some residual dark symmetry.     

Hence, the presence of a dark gauge symmetry under which an electroweak Higgs field is charged can manifestly explain the $W$-boson mass anomaly, and besides the dark charge breaking might further induce neutrino mass and stabilize dark matter. This novel observation can be realized by revisiting a number of compelling gauge extensions of the standard model, presented in order.      

\section{Dark charge as a mirror of hypercharge: a toy model} 

A hidden gauge symmetry $U(1)_M$ that is accidentally conserved by the standard model is usually investigated in the literature in which $M$ may refer to the baryon number minus lepton number $B-L$ \cite{Davidson:1978pm,Mohapatra:1980qe,Marshak:1979fm}, a difference of lepton flavors $L_i-L_j$ for $i,j=e,\mu,\tau$ \cite{Foot:1990mn,Foot:1994vd,He:1991qd}, or a $\mathcal{X}$-charge that vanishes for normal matter \cite{Holdom:1985ag}. Since both the hypercharge $Y$ and the hidden charge $M$ are conserved, a combination of type, \be Y' = Y-\delta M,\label{eq5}\ee is conserved too. In this way, we call $Y'$ to be a mirror of $Y$, transformed by $M$ through a $\delta$ parameter. Because of $Q=T_3+Y$, we find a new charge, \be Q'\equiv T_3+Y'=Q-\delta M,\label{eq2}\ee to be a mirror of $Q$ by the same $M$ transformation. $Q'$ (thus $Y'$) may be regarded as a dequantization effect of electric charge in the standard model, as derived in \cite{VanDong:2020cjf,VanLoi:2020kdk,VanLoi:2021dzv}. Hence, we propose a full gauge symmetry, 
\be SU(3)_C\otimes SU(2)_L\otimes U(1)_Y\otimes U(1)_{Y'},\label{eq1}\ee where $Y,Y'$ determine $Q,Q'$ through the same $T_3$ operator, respectively. Although the Lagrangian conserves both $U(1)_M$ and $U(1)_{Y'}$, which differ only by a $Y$ transformation, the weak vacuum does not conserve $Y'$ in contrast to $M$ as often studied. The dark charge $Y'$ broken by the weak vacuum would cause a $Z$-boson mass shift, enhancing the $\rho$-parameter, thus $T$-parameter, explaining the $W$-boson mass deviation, as shown below (see also \cite{Strumia:2022qkt}).  

\begin{table}[h]
\begin{tabular}{lcccc}
\\
\hline\hline
Multiplet & $SU(3)_C$ & $SU(2)_L$ & $U(1)_{Y}$ & $U(1)_{Y'}$\\
\hline
$l_{L}=\begin{pmatrix}
\nu_{L}\\
e_{L}
\end{pmatrix}$ & 1 & 2 & $-1/2$ & $-1/2+\delta$\\
$\nu_{R}$ & 1 & 1& 0 & $\delta$ \\
$e_{R}$ & 1 & 1 & $-1$ & $-1+\delta$\\
$q_{L}=\begin{pmatrix}
u_{L}\\
d_{L}
\end{pmatrix}$ & 3 & 2 & 1/6 & $1/6-\delta/3$ \\
$u_{R}$ & 3 & 1 & 2/3 & $2/3-\delta/3$\\
$d_{R}$ & 3 & 1 & $-1/3$ & $-1/3-\delta/3$\\
$H=\begin{pmatrix}
H^+\\
H^0
\end{pmatrix}$ & 1 & 2 & 1/2 & 1/2\\
$\chi$ & 1 & 1& 0 & $-2\delta$ \\
\hline\hline
\end{tabular} 
\caption[]{\label{tab1} Field representation with $M=B-L$.}
\end{table}

We take $M=B-L$ into account, while the other cases of $M$ can be straightforwardly generalized. We impose $\nu_{R}$ each for a family for anomaly cancellation relevant to $Y'=Y-\delta(B-L)$ and a scalar singlet $\chi$ that couples to $\nu_{R}\nu_R$, breaking $Y'$. The particle content according to the gauge symmetry is given in Table \ref{tab1}. The scalars $\chi,H$ develop a vacuum expectation value (vev),
\be \langle\chi\rangle=\Lambda/\sqrt{2},\hs \langle H \rangle =\begin{pmatrix}
0\\
v/\sqrt{2}\end{pmatrix},\ee where we impose $v=v_{\mathrm{w}}\ll \La$ for consistency. $\Lambda$ breaks $Y'$, providing a Majorana mass for $\nu_R\nu_R$, while $v$ breaks weak isospin and both $Y,Y'$, supplying a Dirac mass for $\nu_L\nu_R$. Hence, the observed neutrino mass is suitably induced, $m_\nu\sim v^2/\La$, by a canonical seesaw, because of $\La\gg v$. Although $v$ conserves both $Q,Q'$, the vev $\La$ breaks $Q'$ (while leaving $Q$ conserved) to a discrete residual symmetry $R=(-1)^{k Q'}$ for $k$ integer, responsible for dark matter stability, as shown in \cite{VanDong:2020cjf,VanLoi:2020kdk,VanLoi:2021dzv}. 

Apart from the QCD, the covariant derivative takes the form, $\mathcal{D}_\mu =\pa_\mu + i g T_j A_{j\mu} + i g_{B} Y B_\mu + i g_{B'} Y' B'_\mu$, where $A_{j}$ $(j=1,2,3)$ and $B$ define the usual gauge fields, $W^\pm$, $Z$, and $\ga$, while $B'$ is a new gauge boson. Additionally, $g$, $g_B$, and $g_{B'}$ are $SU(2)_L$, $U(1)_Y$, and $U(1)_{Y'}$ couplings, respectively. After the symmetry breaking by $v,\La$, the model reveals a $Z$-$B'$ mixing, reducing the $Z$ boson mass, thus a nonzero $T$-parameter at tree level as 
\be \al T=\rho-1\simeq \fr{v^2}{16\delta^2\La^2}.\ee This matches the proposal (\ref{eq}) because of $Y'(H)= 1/2$ and $m_{B'}=2|\delta|g_{B'} \La$. The CDF $W$-boson mass limits $\delta\La\simeq 1.8$ TeV. Further, the $Z$-neutrino coupling is modified, shifting the invisible $Z$ width, $\Delta \Ga_{\mathrm{inv}}/\Ga_{\mathrm{inv}}\simeq v^2/8\delta^2\La^2$. Comparing with $\Delta \Ga_{\mathrm{inv}}/\Ga_{\mathrm{inv}}\lesssim 0.005$ \cite{pdg} gives $\delta \La\gtrsim 1.23$ TeV. The interaction of $B'$ with fermions violates parity symmetry, contributing to Cesium weak charge, $\Delta Q_W(\mathrm{Cs})\simeq 18v^2/\delta^2\La^2$. Comparing with $\Delta Q_W(\mathrm{Cs})<0.61$ \cite{pdg} yields $\delta \La>1.33$ TeV. The LEPII \cite{ALEPH:2013dgf} and LHC \cite{ATLAS:2017} looked for dilepton signals from $B'$ decay, limiting $m_{B'}/g_{B'}$ at a few TeV, where the exact bound depends on $\delta$, which implies $\delta \La>\mathcal{O}(1)$ TeV. 

Therefore, the relation (\ref{eq5}) presents a mirror of hypercharge viably at TeV scale, solving the $W$-boson mass deviation and the neutrino mass, as well as providing a stability mechanism $R$ for dark matter. However, since every new field of the model transforms trivially under $R$, the dark matter candidates must be included by hand, based upon the stability symmetry, $R$ \cite{VanDong:2020cjf,VanLoi:2020kdk,VanLoi:2021dzv}. Can we have a dark charge scheme manifestly stabilizing dark matter by itself, besides solving the previous questions? In what follow, we will search for a mirror of the weak isospin. 
 
\section{Dark charge as a mirror of weak isospin: a realistic model}

Since every non-abelian algebra is already fixed, the mirror of weak isospin, \be T_j\rightarrow T'_j,\label{eqt8} \ee should yield an independent algebra, which performs $SU(2)'_L$, where a prime indicates the dark side of the normal one \cite{darksu2}. Of course, all the standard model fields transform as $SU(2)'_L$ singlets. In spite of the distinction, the dark side is connected to the normal sector via a second Higgs field, say $\Phi\sim (2,m)$, transforming nontrivially under $SU(2)_L\otimes SU(2)'_L$ for $m=2,3,4,\cdots$, satisfying the criteria (\ref{eq}); that is, the $SU(2)_L$ doublet $\Phi$ should have a nontrivial $SU(2)'_L$ charge as set by $m$ in order to couple to the relevant $SU(2)'_L$ dark gauge boson that consequently induces an effective interaction like (\ref{eq}). The choice of such $m$-dimensional representation for $\Phi$ under $SU(2)'_L$ should appropriately stabilize dark matter. A chiral fermion $\xi$ that is a pure $SU(2)'_L$ $m$-plet and imposed each for a family couples to $\bar{l}_L \Phi$, i.e. \be \mathcal{L}\supset h \bar{l}_L\Phi \xi - (1/2)M \xi\xi+ H.c.,\label{dd1} \ee responsible for neutrino mass generation.\footnote{The gauge anomalies associated with $\xi$ always vanish, for instance $\mathrm{Tr}[\{T'_j,T'_k\}T'_l]$=0 for any $SU(2)'_L$ representation.} Here $h$ is a dimensionless coupling, $M$ is a mass parameter, and $\xi\xi$ is viable for odd-$m$ but vanishes for even-$m$. Verify Appendix \ref{app} for $SU(2)$ tensor product techniques we introduce for building Lagrangian terms. All the fields are collected in Table \ref{tab2} according to the new gauge group, \be SU(3)_C\otimes SU(2)_L\otimes U(1)_Y\otimes SU(2)'_L.\ee 

\begin{table}[h]
\begin{tabular}{lcccc}
\\
\hline\hline
Multiplet & $SU(3)_C$ & $SU(2)_L$ & $U(1)_{Y}$ & $SU(2)'_L$\\
\hline
$l_{L}=\begin{pmatrix}
\nu_{L}\\
e_{L}
\end{pmatrix}$ & 1 & 2 & $-1/2$ & 1\\
$e_{R}$ & 1 & 1 & $-1$ & 1\\
$q_{L}=\begin{pmatrix}
u_{L}\\
d_{L}
\end{pmatrix}$ & 3 & 2 & 1/6 & 1 \\
$u_{R}$ & 3 & 1 & 2/3 & 1\\
$d_{R}$ & 3 & 1 & $-1/3$ & 1\\
$H=\begin{pmatrix}
H^{+}\\
H^{0}
\end{pmatrix}$ & 1 & 2 & 1/2 & 1\\
$\Phi=\begin{pmatrix}
\Phi^{0}_1 & \Phi^{0}_2 & \cdots & \Phi^{0}_m  \\
\Phi^{-}_1 & \Phi^{-}_2 & \cdots & \Phi^{-}_m
\end{pmatrix}$ & 1 & 2 & $-1/2$ & $m$\\
$\xi=\begin{pmatrix}
\xi_{1}\\
\xi_{2}\\
\vdots\\
\xi_{m} 
\end{pmatrix} $ & 1 & 1 & 0 & $m$ \\
$\varphi=\begin{pmatrix}
\varphi_1\\
\varphi_2\\
\vdots \\
\varphi_n
\end{pmatrix} $ & 1 & 1 & 0 & $n$ \\
\hline\hline
\end{tabular} 
\caption[]{\label{tab2} Field representation with isospin mirror.}
\end{table}

A scalar multiplet, called $\varphi=(\varphi_1,\varphi_2,\varphi_3,\cdots,\varphi_n)$, as in the table is needed for $SU(2)'_L$ breaking, governed by a potential $\mu^2_\varphi (\varphi^\dagger \varphi)+\la_\varphi (\varphi^\dagger \varphi)^2$ for $\mu^2_\varphi<0$ and $\la_\varphi>0$. The vev of $\varphi$ obeys $\langle \varphi\rangle^2=-\mu^2_\varphi/2\la_\varphi\equiv \La^2/2$ and we choose the vacuum alignment $\langle \varphi \rangle = (\La/\sqrt{2},0,0,\cdots,0)$, where $\varphi_1$ has the highest weight $T'_3=(n-1)/2$. All the generators of $SU(2)'_L$ are broken by $\La$, but it may preserve a residual symmetry, satisfying $e^{i\al_j T'_j}\langle \varphi\rangle = \langle \varphi\rangle$. We derive $\al_1=\al_2=0$, while $\al_3=k4\pi/(n-1)$ for $k$ integer. Hence, the residual symmetry is $P=e^{i\fr{k4\pi}{n-1}T'_3}$. We rewrite $P=p^{k2T'_3}=D^k$, where $p \equiv e^{i\fr{2\pi}{n-1}}$ is the $(n-1)$th root of unity, while $D \equiv p^{2T'_3}$ is the generator of $P$. Since $2T'_3$ is integer, $D^{n-1}=1^{2T'_3}=1$. The residual symmetry is automorphic to $Z_{n-1}$, i.e. 
\be P=\{1,D,D^2,\cdots,D^{n-2}\}\cong Z_{n-1}.\ee Exceptionally, if there are only odd-dimensional $SU(2)'_L$ representations present in the model, $T'_3$ is integer. The residual symmetry is automorphic to \be P=\{1,D,D^2,\cdots, D^{(n-3)/2}\}\cong Z_{(n-1)/2},\label{dd12}\ee since $D^{(n-1)/2}=1^{T'_3}=1$.  

After $SU(2)'_L$ breaking by $\langle \varphi\rangle$, the electroweak symmetry is broken by the Higgs vacuum, $\langle H\rangle = (0,v/\sqrt{2})$, also by $\langle \Phi\rangle$ that appropriately conserves $P$. We impose $\La\gg v, \langle \Phi\rangle$ for consistency with the standard model. The scheme of symmetry breaking is summarized as  
\bc \begin{tabular}{c} $SU(2)_L\otimes U(1)_Y\otimes SU(2)'_L$ \\
$\downarrow\langle \varphi\rangle $\\
$SU(2)_L\otimes U(1)_Y\otimes P$\\
$\downarrow \langle H,\Phi\rangle $\\
$U(1)_Q\otimes P$
\end{tabular}\ec  The electric charge takes the usual form, $Q=T_3+Y$. Since $\Phi, \varphi$ do not couple to ordinary fermions like $H$ in Yukawa interactions, the charged fermions gain suitable masses only from the vev of $H$, similar to the standard model. We will not refer to this matter further. 

The model classes with respect to values of $n$ will be investigated in order, in which within each model class specific versions emerged dependently on $m$ are signified. Before proceeding further, let us call the reader's attention to \cite{hambye,nomura,gross,baek,khoze,nomura1} for particular realizations of dark matter stability and to \cite{wmasssu2,wmasssu2u1} for alternative explanations of the $W$-mass shift, concerning a dark isospin. 
 
\subsection{The model class with $n=2$} 

In this case, $\varphi$ is a doublet similar to that of the standard model. The residual symmetry becomes $P=1$ which is trivial. This model class by itself does not stabilize dark matter for any $m$, thus not favored.\footnote{Intriguingly, this points out that the proposal in \cite{hambye} is unique.} 

\subsection{The model class with $n=3$} 

The field $\varphi$ is a triplet, $\varphi=(\varphi_1,\varphi_2,\varphi_3)$, with vev $\langle \varphi\rangle = (\La/\sqrt{2},0,0)$. The residual symmetry is \be P=(-1)^{k2T'_3}=\{1,(-1)^{2T'_3}\}\cong Z_2,\ee i.e. every (odd-) even-dimensional representation is $Z_2$-odd (-even). This model class does not allow dark matter resided in dark gauge boson, $\varphi$, even $\Phi,\xi$ for odd-$m$. 

The simplest possibility is $m=2$ by which both $\Phi,\xi$ are $Z_2$-odd responsible for dark matter, realizing the scotogenic setup, because of the coupling $\bar{l}_L\Phi \xi$ in (\ref{dd1}) \cite{scoto}. 

Majorana neutrino mass is generated by one-loop contribution of both $\Phi$ and $\xi$ in the loop, where the Majorana $\xi$ mass is derived by $\xi\xi\varphi$ coupling, while the mass splitting of real and imaginary $\Phi^0$ parts is given by an effective interaction, $(H\Phi)^2\varphi/\La$. The $W$-boson mass anomaly is explained by $T_{\mathrm{rad}}$ due to $\Phi$ contribution. 

Notice that the coupling $(H\Phi)^2$ vanishes as $\xi\xi$ does, which requires the effective interaction for neutrino mass generation. The status remains unchanged for $m=4,6,\cdots$. Hence, the scotogenic setup is only realized at level of non-renormalizable interactions, which is not of interest in this work. In the following, we interpret only renormalizable couplings too.        

\subsection{The model class with $n=4$} 

The field $\varphi$ is a quartet, namely $\varphi=(\varphi_1,\varphi_2,\varphi_3,\varphi_4)$, having the vev $\langle \varphi\rangle=(\La/\sqrt{2},0,0,0)$. The residual symmetry is \be P=w^{k2T'_3}=\{1,w^{2 T'_3},w^{4T'_3}\}\cong Z_3,\ee where $w=e^{i2\pi/3}$ is the cube root of unity, and the generator is $D=w^{2T'_3}$. The relevant fields gain $D$ values, such as \ben \item $D\varphi=(1,w,w^2,1)$ for $\varphi$, \item  $DA'=(w^2,1,w)$ for dark gauge boson $A'=(A'^+,A'^0,A'^-)$ arranged in $T'_3$ weight order, \item $D\Phi,D\xi=(w,w^2)$ for $m=2$. Otherwise, $D\Phi,D\xi$ are similar to $A'$ for $m=3$ and $\varphi$ for $m=4$ as $n$.\een This class of model provides dark matter candidates to be $\varphi_{2,3}$, $A'^{\pm}$, and those resided in $\Phi,\xi$ depending on $m$. 

The model with $m=2$ cannot induce neutrino mass, since $\langle \Phi\rangle =0$ by $Z_3$ conservation and that the $\xi\xi$ mass vanishes.

The model with $m=4$ yields $D\Phi,D\xi=(1,w,w^2,1)$ for $\Phi=(\Phi_1,\Phi_2,\Phi_3,\Phi_4)$ and $\xi=(\xi_1,\xi_2,\xi_3,\xi_4)$. Hereafter, each $\Phi_i$ denotes an electroweak doublet $(\Phi^0_i,\Phi^-_i)$ as explicitly set in Table \ref{tab2}. The $Z_3$ symmetry suppresses the vev of $\Phi_{2,3}$, but the remainders can develop a vev, 
\be \langle \Phi\rangle =\begin{pmatrix}
u_1/\sqrt{2} & 0 & 0 & u_4/\sqrt{2}\\
0 & 0 & 0 & 0
\end{pmatrix},\ee where $u_{1,4}$ are at the weak scale, $\sqrt{v^2+u^2_1+u^2_4}=v_{\mathrm{w}}$. The neutrino mass generation Lagrangian has a simple form $\mathcal{L}\supset h\bar{l}_L \Phi \xi +H.c.$ since $\xi\xi$ mass vanishes. Neutrino obtains a Dirac mass, $\mathcal{L}\supset \fr{h}{2\sqrt{2}}\bar{\nu}(u_1\xi_4-u_4\xi_1)$, once combined with the state $u_1\xi_4-u_4\xi_1$. The remaining states $u_1\xi_1+u_4\xi_4$ and $\xi_{2,3}$ are massless. Since the induced Dirac neutrino mass $\sim u_{1,4}$ is not naturally small, this case is not favored, similarly to $m=2$.       
 
The model with $m=3$ yields $D\Phi,D\xi=(w^2,1,w)$ for $\Phi=(\Phi_1,\Phi_2,\Phi_3)$ and $\xi=(\xi_1,\xi_2,\xi_3)$. The Yukawa Lagrangian relevant to $\nu,\xi$ is $\mathcal{L}\supset h\bar{l}_L\Phi \xi -(1/2) M \xi\xi +H.c.$ as in (\ref{dd1}). Only $\Phi^0_2$ in $\Phi$ can develop a vev conserving both $Z_3$ and electric charge, i.e. \be \langle \Phi\rangle =\begin{pmatrix}
0 & u/\sqrt{2} & 0\\
0 & 0 & 0\end{pmatrix},\ee where $u$ is a weak scale obeying $\sqrt{v^2+u^2}=v_{\mathrm{w}}$. Hence, the $\nu$-$\xi$ mass terms are induced as \be\mathcal{L}\supset -\fr 1 2 (\bar{\nu}_L\ \bar{\xi}^c_2)\begin{pmatrix}
0 & \sqrt{\fr 2 3} u h\\
 \sqrt{\fr 2 3} u h&-\fr{M}{\sqrt{3}} \end{pmatrix}\begin{pmatrix} \nu^c_L\\
\xi_2 \end{pmatrix}+H.c.\label{dd123} \ee  Assuming $u\ll M$, the neutrino gains a small mass, $m_\nu\simeq 2h^2 u^2/\sqrt{3}M$, taking the form of canonical seesaw in $SU(2)_L$ sense, but mirrored to a type III seesaw on the $SU(2)'_L$ dark side. The mass of $\xi_2$ is $m_{\xi_2}\simeq -M/\sqrt{3}$. Two $Z_3$ fermions $\xi_{1,3}$ do not mix with $\xi_2$, having a degenerate mass $\pm M/\sqrt{3}$. It is noted that $A'^\pm$ do not mix with $Z$ boson, by $Z_3$ conservation. Additionally, $A'^0$ does not mix with $Z$ too, since $\Phi^0_2$ has zero $T'_3$ charge. Hence, $W,Z$ bosons are physical fields by themselves with mass $m_W=gv_{\mathrm{w}}/2$ and $m_Z=gv_{\mathrm{w}}/2c_W$, implying $T_{\mathrm{tree}}=0$. The $W$-boson mass anomaly is solved by $T_{\mathrm{rad}}$ by $\Phi$ contribution, similar to the model with $n=3$. 

\subsection{The model class with $n=5$} 

Lastly, we consider the case with $n=5$, thus $\varphi$ is a quintet, $\varphi=(\varphi_1,\varphi_2,\varphi_3,\varphi_4,\varphi_5)$, possessing vev $\langle \varphi\rangle =(\La/\sqrt{2},0,0,0,0)$. The residual symmetry is \be P=e^{ik\pi T'_3}=\{1,D,D^2,D^3\}\cong Z_4,\ee where $D= e^{i\pi T'_3}=(-1)^{T'_3}$. The fields that have nontrivial $D$ are obtained as
\ben
\item $D\varphi = (+,-,+,-,+)$ for $\varphi$,
\item $DA'=(-,+,-)$ for $A'=(A'^+,A'^0,A'^-)$,
\item $D\Phi,D\xi=(i,-i)$ for $m=2$. Alternatively, $D\Phi,D\xi =(-i,i,-i, i)$ for $m=4$, and $D\Phi,D\xi$ are analogous to $A'$ for $m=3$ and $\varphi$ for $m=5$ as $n$.
\een The dark matter candidates include $\varphi_{2,4}$, $A'^\pm$, and those resided in $X=\Phi,\xi$ dependent on $m$. 

The model with $m=2,4$ does not give neutrino mass, since $\langle \Phi\rangle =0$ and $\xi\xi$ mass vanishes, similar to a previous model (note that $\xi\xi \varphi$ vanishes for $m=4$). 

The model with $m=3$ gives $D\Phi,D\xi=(-,+,-)$ for $\Phi=(\Phi_1,\Phi_2,\Phi_3)$ and $\xi=(\xi_1,\xi_2,\xi_3)$. Since this model contains only odd-dimensional $SU(2)'_L$ representations, the residual symmetry is actually $P=\{1,D\}\cong Z_2$ due to $D^2=1$, as mentioned in (\ref{dd12}), in agreement with \cite{nomura}. The Lagrangian relevant to $\nu,\xi$ mass is \be \mathcal{L}\supset h\bar{l}_L\Phi\xi -(1/2) (M+f\varphi) \xi\xi +H.c.,\label{dd1234}\ee where $\varphi$ has a coupling to $\xi\xi$, thus splitting $\xi$ mass responsible for dark matter. The electric charge and $Z_2$ conservation allows only $\Phi_2^0$ in $\Phi$ to have a vev, $\langle \Phi^0_2\rangle = u/\sqrt{2}$, and $\sqrt{v^2+u^2}=v_{\mathrm{w}}$, as usual.  Substituting the vevs of $\Phi,\varphi$ to (\ref{dd1234}), we obtain the neutrino mass matrix identical to (\ref{dd123}), i.e. the $\nu,\xi_2$ masses are $m_\nu\simeq 2h^2 u^2/\sqrt{3}M$ and $m_{\xi_2}\simeq -M/\sqrt{3}$, for $u\ll M$. The odd fields $\xi_{1,3}$ do not mix with $\xi_2$, having masses heavily separated by $\La$, unlike the above case. The usual gauge bosons do not mix with $A'$, obtaining tree-level masses, $m_{W}=gv_{\mathrm{w}}/2$ and $m_{Z} = gv_{\mathrm{w}}/2c_W$, thus $T_{\mathrm{tree}}=0$. Hence, the $W$-boson mass anomaly arises only from radiative $\Phi$ contribution.     

Last, but not least, the model with $m=5$ leads to $\Phi=(\Phi_1,\Phi_2,\Phi_3,\Phi_4,\Phi_5)$ and $\xi=(\xi_1,\xi_2,\xi_3,\xi_4,\xi_5)$ with $D\Phi,D\xi=(+,-,+,-,+)$, which are like the size and parity of $\varphi$. The residual symmetry is $P=\{1,D\}\cong Z_2$, similar to $m=3$. The field $\Phi$ can develop a vev, 
\bea 
\langle \Phi\rangle &=& \begin{pmatrix}
u_1/\sqrt{2} & 0 & u_3/\sqrt{2} & 0 & u_5/\sqrt{2} \\
0 & 0 & 0 & 0& 0
\end{pmatrix}, \eea satisfying $\sqrt{u^2_1+u^2_3+u^2_5+v^2}=v_{\mathrm{w}}$. The $\nu,\xi$ mass generation Lagrangian is identical to (\ref{dd1234}), thus we obtain the relevant mass terms,
\bea && -\fr 1 2 (\bar{\nu}_L\ \bar{\xi}^c_1\ \bar{\xi}^c_3\ \bar{\xi}^c_5)\begin{pmatrix}
0 & -\fr{hu_5}{\sqrt{10}} & -\fr{hu_3}{\sqrt{10}} & -\fr{hu_1}{\sqrt{10}}\\
-\fr{hu_5}{\sqrt{10}} & 0 & 0 & \fr{M}{\sqrt{5}}\\
-\fr{hu_3}{\sqrt{10}} & 0 & \fr{M}{\sqrt{5}} & \fr{f\La}{\sqrt{35}}\\
-\fr{hu_1}{\sqrt{10}} & \fr{M}{\sqrt{5}} & \fr{f\La}{\sqrt{35}} & 0
 \end{pmatrix}\begin{pmatrix} \nu^c_L\\
\xi_1\\
\xi_3\\
\xi_5 \end{pmatrix}\crn
&& -\fr 1 2  (\xi_2\ \xi_4)\begin{pmatrix}
0 & -\fr{M}{\sqrt{5}}\\
-\fr{M}{\sqrt{5}} & -\sqrt{\fr{3}{70}} f\La\end{pmatrix}
\begin{pmatrix}
\xi_2\\
\xi_4\end{pmatrix}+ H.c.\label{edtt1} \eea
Given that $u_{1},u_3,u_5\ll M,\La$, the neutrino gains a naturally small mass, $m_\nu\sim h^2(u_1,u_3,u_5)^2/(M,\La)$, via the seesaw mechanism. The fields $\xi$'s are heavy at $M,\La$ scale and completely separated. Let $g'$ be $SU(2)'_L$ gauge coupling. We obtain nonzero gauge boson mass terms,\bde \bea && m^2_W W^+ W^-+\fr 1 2 (Z\ A'^0)
\begin{pmatrix}
m^2_Z & \fr{gg'}{c_W}(u^2_1-u^2_5)\\
\fr{gg'}{c_W}(u^2_1-u^2_5) & m^2_{A'^0}\end{pmatrix}
\begin{pmatrix}
Z\\
A'^0
\end{pmatrix}\crn
&&+\fr 1 2 (A'^+\ A'^-)\begin{pmatrix}
\sqrt{6}g'^2u_3(u_1+u_5)& m^2_{A'^\pm}\\
m^2_{A'^\pm}& \sqrt{6}g'^2u_3(u_1+u_5) \end{pmatrix} 
\begin{pmatrix}
A'^+\\
A'^-\end{pmatrix},\eea\ede where we define $m_W=gv_{\mathrm{w}}/2$, $m_Z=gv_{\mathrm{w}}/2c_W$, $m^2_{A'^0}=4g'^2(\La^2+u^2_1+u^2_5)$, and $m^2_{A'^\pm}=g'^2(\La^2+u^2_1+u^2_5+3u^2_3)$. The $Z$ boson mixes with $A'^0$ due to the vev $u_{1,5}$, shifting its mass by an amount $\Delta m^2_{Z}\simeq -g^2g'^2(u^2_1-u^2_5)^2/c^2_Wm^2_{A'^0}$ by the seesaw formula, since $u_{1,3,5}\ll \La$. Therefore, it contributes to the $T$-parameter as
\be \al T_{\mathrm{tree}}=\rho-1\simeq \fr{-\Delta m^2_Z}{m^2_Z}\simeq \fr{(u^2_1-u^2_5)^2}{v^2_{\mathrm{w}}\La^2}. \ee This matches the prediction (\ref{eq}) for which the above result is induced by both Higgs doublets $\Phi_{1,5}$ coupled to the dark gauge boson, such as $\mathcal{L}\supset -i2g'(\Phi^\dagger_1 D^\mu\Phi_1-\Phi^\dagger_5 D^\mu \Phi_5)A'^0_\mu+H.c.$ The $W$-boson mass deviation gives $\La\simeq 30 |u^2_1-u^2_5|/v_{\mathrm{w}}\sim 7$ TeV, given that $u_{1,5}\sim v_{\mathrm{w}}$. Finally, the new gauge bosons $A'^0$, $A'^+$, and $A'^-$ all obtain masses at $\La$ scale, completely separated by $u_{1,3,5}$. 

\section{Dark matter observables}

Among the mentioned models, we choose the one with $n=m=5$ which solves the neutrino mass and the $W$-boson mass deviation naturally at tree-level. The Lagrangian of this model and necessary ingredients are supplied in Appendix \ref{appb}.

This model contains dark fields, $\xi_{2,4}$, $\Phi^0_{2,4}$, $\varphi_{2,4}$, and $A'^\pm$, which are $Z_2$-odd. The fields $\Phi^0_{2,4}$ and $\xi_{2,4}$ interact with normal fields directly via $\mathcal{L}\supset h\bar{l}_L\Phi \xi$ and/or the usual gauge portal, which are relevant to the neutrino mass and $W$-mass shift, besides the Higgs portals as the other dark fields do. Therefore, it is worth to interpret the dark matter candidate to be of either $\Phi^0_{2,4}$ or $\xi_{2,4}$, whereas the rest of dark fields is all heavier than them. Since $\xi$ has three flavors similar to the lepton $l$, we consider only the lightest flavor of $\xi$, decoupled from the others. 

\subsection{Fermion dark matter}

As given in (\ref{edtt1}), the $\xi_{2,4}$ masses are separated. Diagonalizing the relevant mass matrix, we obtain physical eigenstates, \be \xi_{24}\equiv s_\theta \xi_2 + c_\theta \xi_4,\hs \xi'_{24}\equiv c_\theta \xi_2 -s_\theta \xi_4,\ee defined via a mixing angle, $t_{2\theta}=2\sqrt{14/3}M/f\La$, which is not small due to $M\sim \La$. The mass eigenvalues are given by \bea m_{\xi_{24}} &=& -\fr 1 2 \sqrt{\fr{3}{70}}f\La +\fr 1 2 \sqrt{\fr{3}{70}f^2\La^2+\fr 4 5 M^2},\\
m_{\xi'_{24}} &=& -\fr 1 2 \sqrt{\fr{3}{70}}f\La -\fr 1 2 \sqrt{\fr{3}{70}f^2\La^2+\fr 4 5 M^2}.\label{dttn12} \eea Since $\xi_{24}$ is lighter than $\xi'_{24}$, i.e. $|m_{\xi_{24}}|<|m_{\xi'_{24}}|$ for $f>0$, we assume $\xi_{24}$ to be the lightest of all dark fields. Hence, $\xi_{24}$ is stabilized by $Z_2$, responsible for dark matter. 

\begin{figure}[h]
\begin{center}
\includegraphics[]{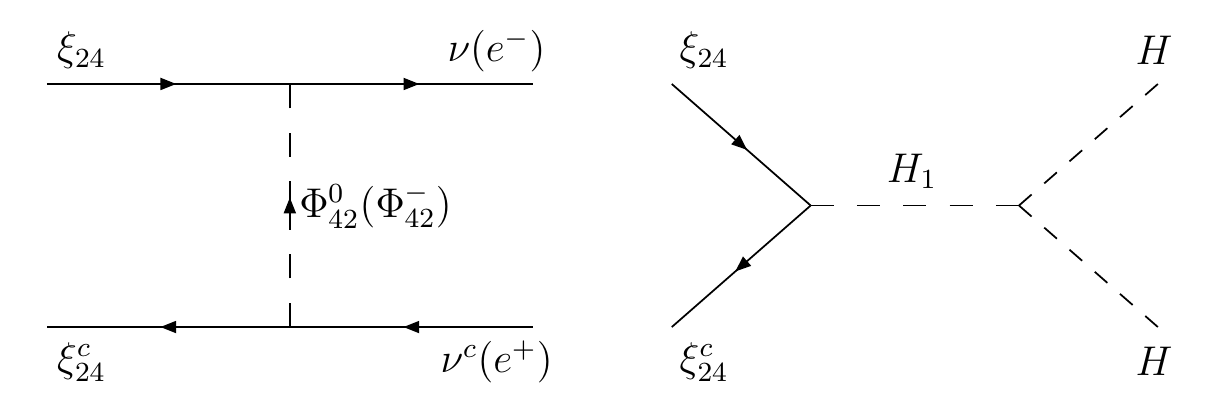}
\caption[]{\label{fig1} Annihilation of fermion dark matter.}
\end{center}
\end{figure}

The annihilation process of $\xi_{24}$ dark matter is described by the diagrams in Figure \ref{fig1}, where the scalar combination $\Phi_{42}\equiv s_\theta \Phi_4 + c_\theta \Phi_2$ couples to $\xi_{24}$ and $l$ via $h$-coupling, while the new Higgs field $H_1$ relevant to the $SU(2)'_L$ breaking, i.e. $\varphi_{1}\rightarrow (\La+H_1)/\sqrt{2}$, couples to $\xi^2_{24}$ via the coupling $\fr 1 2 f c^2_\theta \sqrt{3/70}$ and to the usual Higgs field $H^2$ via the coupling $\fr 1 2 \la_5\La/\sqrt{5}$. 
Applying the Feynman rules, we get the annihilation cross-section,
\be \langle \sigma v\rangle_{\xi_{24}} \simeq \fr{|h|^4 m^2_{\xi_{24}}}{200\pi}\left(\fr{1}{m^4_{\Phi^0_{42}}}+\fr{1}{m^4_{\Phi^-_{42}}}\right)+\fr{3\la^2_5 f^2c^4_{\theta}}{44800\pi}\fr{\La^2}{(4m^2_{\xi_{24}}-m^2_{H_{1}})^2}\sqrt{1-\fr{m^2_H}{m^2_{\xi_{24}}}},\ee since $m_{\nu,e}\ll m_{\xi_{24}}$, and $|h|^2 = \sum_{l,l'}h^*_{l\xi}h_{l' \xi}$ is summed over lepton flavors. 

\begin{figure}[h]
\bc
\includegraphics[scale=0.6]{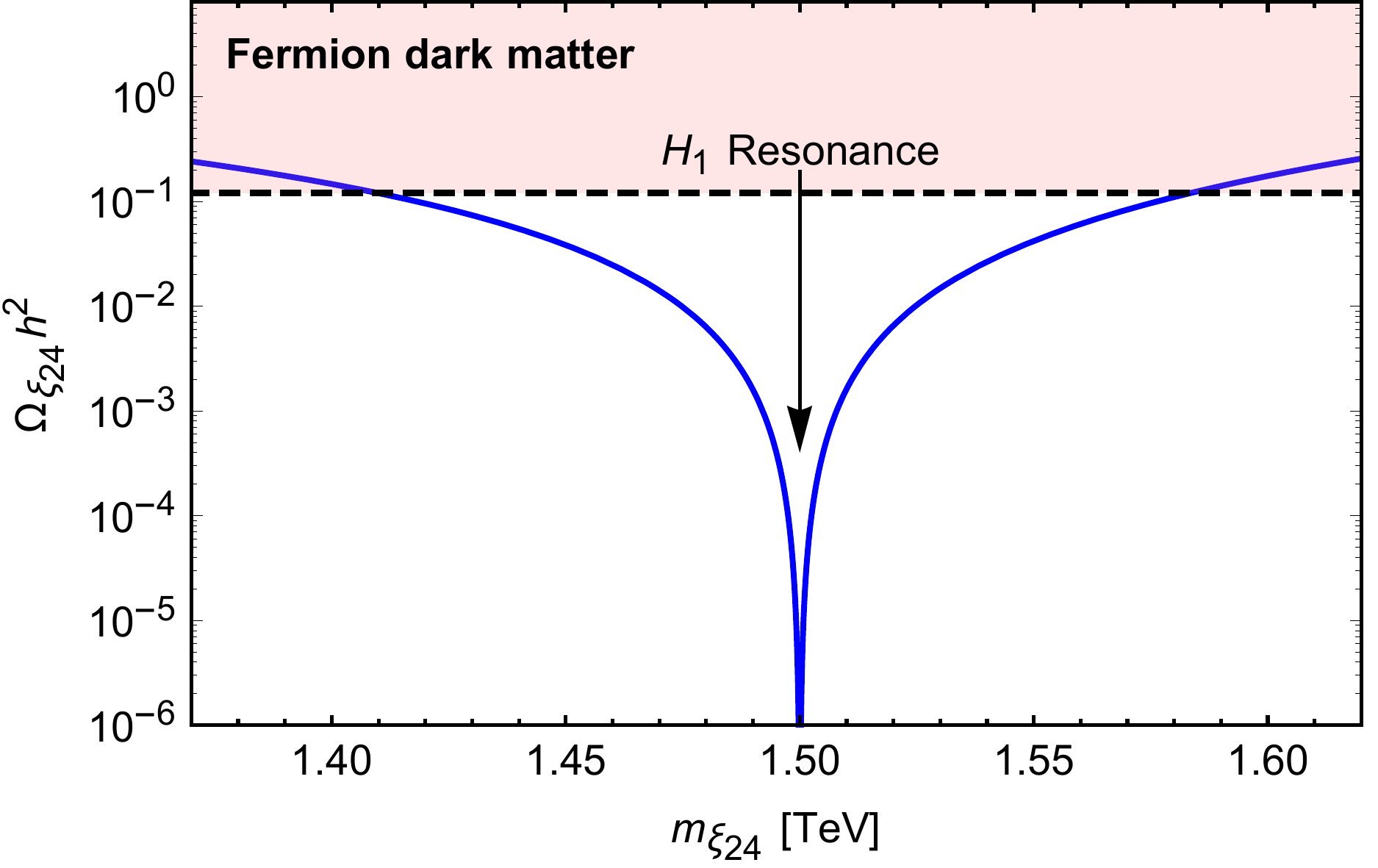}
\caption[]{\label{fig2} Dark matter relic density plotted as the function of its mass.} 
\ec
\end{figure}

The mediator $H_1$ and dark matter $\xi_{24}$ have masses at $\La,M$ scale in TeV regime. Further, the condition for dark matter stability implies that the $\Phi_{42}$ masses are larger than that of $\xi_{24}$. Hence, the $t$-channel processes mediated by $\Phi_{42}$'s contribute negligibly to the annihilation cross-section, because the $h$-coupling required for neutrino mass is analogous to charged lepton couplings, $|h| \sim \sqrt{m_\nu (M,\La)}/v_{\mathrm{w}}\sim 10^{-5}$. In the present model, the $s$-channel process mediated by $H_1$ dominates the annihilation cross-section. Therefore, the dark matter relic density is set by the $H_1$ resonance at which $m_{\xi_{24}} = \fr 1 2 m_{H_1}$, which is at TeV regime, as expected. To be concrete, we plot the relic density $\Om_{\xi_{24}} h^2\simeq 0.1\ \mathrm{pb}/\langle \sigma v\rangle_{\xi_{24}}$ as a function of the dark matter mass as in Figure \ref{fig2}, for a choice of parameters, say $\la_5=1$, $f=2$, $\La=7$ TeV, and $m_{H_1}=3$ TeV. The Higgs mass $m_H=125$ GeV is also used. Note that $\theta$ is related to $m_{\xi_{24}}$ by substituting $M=\{5[(1/2) \sqrt{3/70}f\La+m_{\xi_{24}}]^2-(15/280)f^2\La^2\}^{1/2}$ from (\ref{dttn12}) to $t_{2\theta}=2\sqrt{14/3}M/f\La$. Additionally, since $m_{H_1}= \sqrt{2\la_3}\La$ plus the corrections of other scalar couplings and vevs, fixing its mass correspondingly fixes the relevant scalar couplings, which should be understood. 
 The fermion dark matter gains an appropriate abundance, i.e. $\Om_{\xi_{24}}h^2\leq 0.12$ \cite{pdg}, if $m_{\xi_{24}}=1.41$--1.58 TeV. This corresponds to $M=5.51$--5.96~TeV, as desirable.    

The dark matter $\xi_{24}$ may interact with the usual Higgs field due to a mixing of $H$ with $H_1$. However, the relevant coupling strength is substantially suppressed by $(m_H/m_{H_1})^2$. Hence, the usual Higgs field portal contributes negligibly to the annihilation cross-section, similar to the above $t$-channel. Additionally, in direct detection, the dark matter $\xi_{24}$ scatters with nucleons via the usual Higgs portal, thus suppressed too. Notice that the $H_1$ portal does not contribute to this scattering, since $H_1$ does not interact with quarks and gluons. Furthermore, $\xi_{24}$ may scatter with electrons via $\Phi^-_{42}$ portal, but it gives a tiny recoil energy $2m_e v^2\sim 1$ eV and a small signal strength suppressed by $m_e^2/m^4_{\Phi^-_{42}}$. In other words, $\xi_{24}$ easily escapes every current detection of dark matter.   

\subsection{Scalar dark matter}

\begin{figure}[h]
\begin{center}
\includegraphics[scale=0.8]{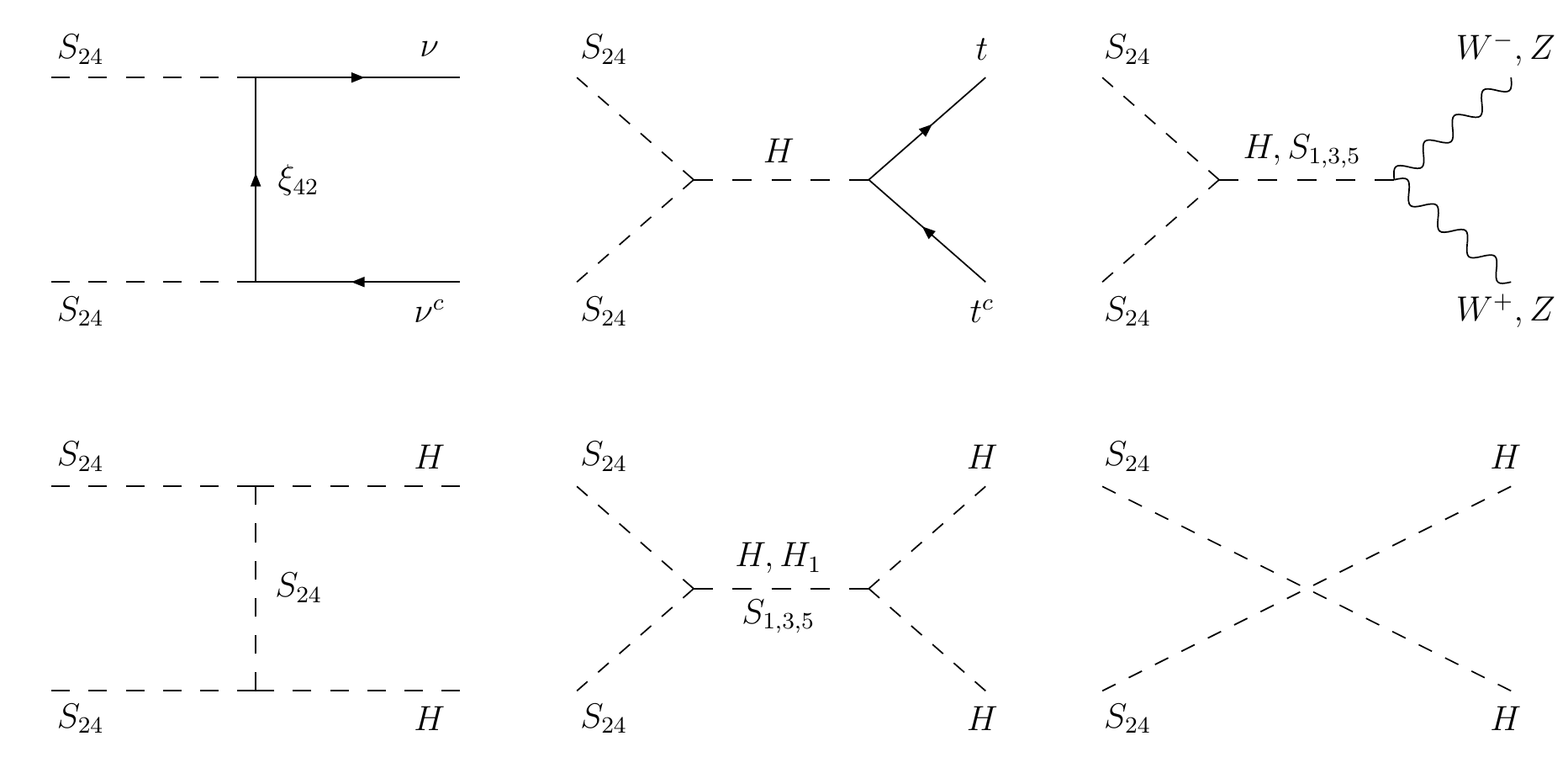}
\caption[]{\label{fig3a} Annihilation of the scalar dark matter via usual/new Higgs and dark fermion portals.}
\end{center}
\end{figure}

As given in Appendix \ref{appb}, let $\Phi^0_2=(S_2+i A_2)/\sqrt{2}$ and $\Phi^0_4=(S_4+i A_4)/\sqrt{2}$. The fields $S_2$ and $A_2$, as well as $S_4$ and $A_4$, are separated in mass proportional to the weak scale because of a CP-violating scalar coupling, $\la_{11}$, while $S_2$ and $S_4$, as well as $A_2$ and $A_4$, are split by the new Higgs vacuum $\La$. Similarly to dark fermion $\xi_{2,4}$, the mixing of $S_2$ $(A_2)$ with $S_4$ $(A_4)$ is arbitrary. Indeed for the current potential, they maximally mix, resulting in a physical state, $S_{24}=(S_2-S_4)/\sqrt{2}$, to be the lightest of all dark fields. It is a dark matter candidate. It annihilates to normal matter via both the usual/new Higgs fields and dark fermion $\xi_{42}=(\xi_4-\xi_2)/\sqrt{2}$ portals, as supplied in Figure \ref{fig3a}. As a weak doublet, $S_{24}$ also annihilates to normal matter via the usual gauge portal, as given in Figure \ref{fig3b}. Here $H_1$ and $S_{1,3,5}$ are specified in Appendix \ref{appb}, while $\Phi^-_{24}$ and $A_{24}$ are weak doublet components that couple to $S_{24}$ through $W$ and $Z$, respectively. 
\begin{figure}[h]
\begin{center}
\includegraphics[scale=0.8]{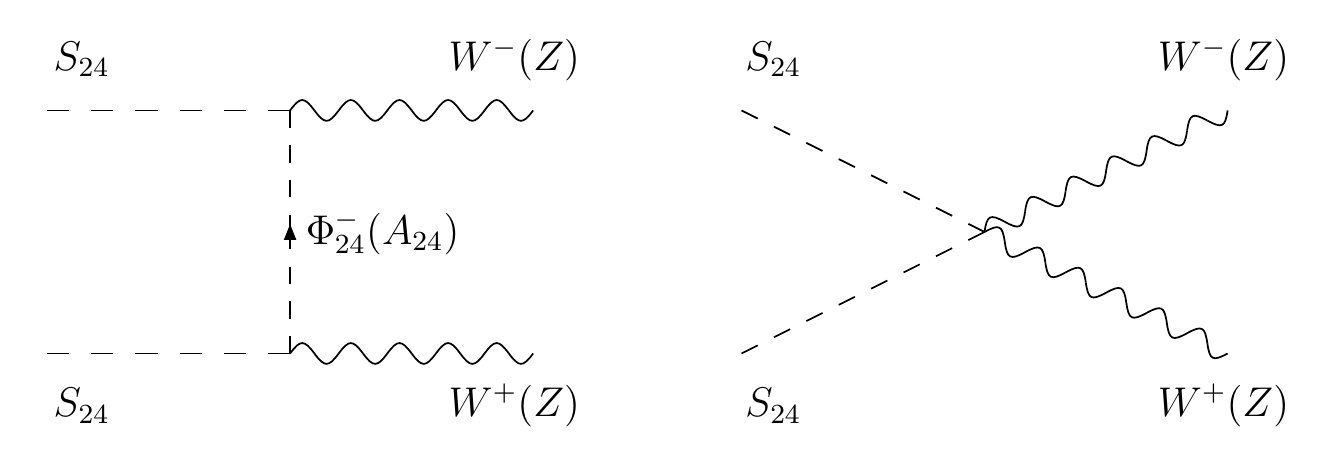}
\caption[]{\label{fig3b} Annihilation of the scalar dark matter through the usual gauge portal.}
\end{center}
\end{figure}
  
We divide into two regimes for the mass of $S_{24}$ dark matter. \ben \item Below TeV: In this regime, $S_{24}$ is lighter than all the new particles. Summarizing all the contributions, the annihilation cross-section is approximated as 
\be \langle \sigma v \rangle_{S_{24}} \simeq 1\ \mathrm{pb}\left[\left(\fr{600\ \mathrm{GeV}}{m_{S_{24}}}\right)^2+ \left(\fr{1.354\bar{\la}\ \mathrm{TeV}}{m_{S_{24}}}\right)^2\right],\ee where the first term comes from the gauge portal, while the last term arises from the usual Higgs portal with $\bar{\la} \equiv (\la_4+\la_{10}-2\la_{11})/4\sqrt{5}$. The dark fermion and new Higgs portals give a negligible contribution. Comparing with the correct abundance, i.e. $\Om_{S_{24}} h^2\simeq 0.1\ \mathrm{pb}/\langle \sigma v\rangle_{S_{24}}\simeq 0.12$, we get $m_{S_{24}} = \sqrt{1+5\bar{\la}^2}\times 600\ \mathrm{GeV}\gtrsim 600$ GeV, dependent on $\bar{\la}$. \item At TeV: In this regime, the new particle resonances govern the dark matter relic density. The relic density gains values appropriate to experiment, given that $m_{S_{24}}$ is around $\fr 1 2 m_{H_1}$ or $\fr 1 2 m_{S_{1,3,5}}$. The phenomenological processes that set the scalar dark matter abundance in this regime are quite similar to the fermion dark matter, and we will not consider it further. \een 

However, differently from the fermion candidate $\xi_{24}$, the scalar dark matter $S_{24}$ can scatter with nucleon ($N$) via the usual Higgs portal in direct detection. The scattering cross-section is given by
\be\sigma^{\mathrm{SI}}_{S_{24}}\simeq \fr{4 m^2_N}{\pi} \la^2_N, \ee where $\la_N$ is the effective coupling of $S_{24}$ with the nucleon, related to that with quarks ($\la_q$), which are confined in $N$, such as 
\be \fr{\la_N}{m_N}=\sum_{u,d,s}f^N_{Tq}\fr{\la_q}{m_q}+\fr{2}{27} f^N_{TG} \sum_{c,b,t} \fr{\la_q}{m_q},\ee where $f^N_{TG}=1-\sum_{u,d,s}f^N_{Tq}$, and $f^N_{Tq}$ takes values given in \cite{johnellis}. The effective dark matter quark coupling takes the form of $\mathcal{L}_{\mathrm{eff}}\supset 2\la_q m_{S_{24}} S_{24} S_{24} \bar{q}q$, mediated by the usual Higgs boson, thus given by \be \la_q=\fr{\bar{\la} m_q}{2m_{S_{24}} m^2_H},\ee where $\bar{\la} = (\la_4+\la_{10}-2\la_{11})/4\sqrt{5}$, as given. For $m_N=1$ GeV and $m_H=125$ GeV, we have 
\be \sigma^{\mathrm{SI}}_{S_{24}}\simeq \bar{\la}^2 \left(\fr{1\ \mathrm{TeV}}{m_{S_{24}}}\right)^2\times 6.15\times 10^{-44}\ \mathrm{cm}^2.\ee It obeys the direct detection bound, $\sigma^{\mathrm{SI}}_{S_{24}}\sim 10^{-45}\ \mathrm{cm}^2$, measured by the XENON experiment \cite{ddetection}, given that $\bar{\la}\sim 0.13$, for the dark matter mass at TeV regime, $m_{S_{24}}\sim 1$ TeV. However, recently the LUX-ZEPLIN collaboration \cite{ddetectionupdate} has provided a stronger constraint, $\sigma^{\mathrm{SI}}_{S_{24}}\sim 10^{-46}\ \mathrm{cm}^2$, for the same dark matter mass regime, i.e. $m_{S_{24}}\sim 1$ TeV, which requires a smaller coupling, $\bar{\la}\sim 0.04$.

\section{Conclusion}

The recently-measured $W$-boson mass anomaly indicates that the Higgs sector of the standard model is perhaps nontrivial, linked to the other questions of new physics. If the usual Higgs doublet has an abelian dark gauge charge, the dark charge breaking induces the $W$-mass deviation and neutrino mass appropriately, but the dark matter candidates must be included by hand. If one introduces a non-abelian dark gauge charge, there are several Higgs doublets charged under this symmetry. Consequently, the symmetry breaking may automatically induce the $W$-mass deviation, neutrino mass, and dark matter stability. 

Considering a dark isospin symmetry, we have shown that the model classes with $n=3,4,5$ possess several versions addressing all the mentioned questions. However, the $W$-mass deviation and neutrino mass are naturally given at tree-level only if $n=m=5$ as the minimal choice. For this case, the neutrino mass is induced by a seesaw mechanism with the contribution of three chiral fermions $\xi_{1,3,5}$, while the other partners $\xi_{2,4}$ provide a dark matter candidate with the relic density set by the dark isospin breaking field. The $W$-mass deviation is contributed by a $Z$-mass shift at tree-level due to the presence of the two Higgs doublets $\Phi_{1,5}$ that couple neutrinos to $\xi_{1,5}$. The other partners $\Phi_{2,4}$ also supply potential candidates for dark matter. Interestingly, the dark gauge bosons are completely separated by a vev of the partner $\Phi_3$.

\section*{Acknowledgement} This research is funded by Vietnam National Foundation for Science and Technology Development (NAFOSTED) under Grant No. 103.01-2019.353. DTH acknowledges the financial support of International Centre of Physics, Institute of Physics, VAST under Grant No. ICP-2023.02.

\section*{Data Availability Statement} No Data associated in the manuscript.

\appendix

\section{\label{app} $SU(2)$ tensor products}

Consider two representations $|x\rangle$ and $|y\rangle$ determined in ket-bases $|a \rangle$ for $a=1,2,3,\cdots,n$ and $|\al \rangle$ for $\al=1,2,3,\cdots, m$, such that $|x\rangle = x_a |a\rangle $ and $|y\rangle = y_\al |\al\rangle$, respectively. We can label each representation according to its dimension and entries as $|x\rangle = \underline{n}(1,2,3,\cdots,n)$ and $|y\rangle = \underline{m}(1,2,3,\cdots,m)$. Their tensor product is $|x\rangle |y\rangle=x_a y_\al |a\rangle |\al\rangle = \underline{n}\otimes \underline{m}(\cdots,a\al,\cdots)$. Notice that in usual notation, $\underline{n}$ [$\underline{m}$] corresponds to the spin-$j=(n-1)/2$ [$j'=(m-1)/2$] representation and the index $a=1,2,3,\cdots,n$ [$\al=1,2,3,\cdots,m$] marks weights $j,j-1,\cdots,-(j-1),-j$ [$j',j'-1,\cdots,-(j'-1),-j'$]---the $T_3$ eigenvalues---whose corresponding eigenstates forming the ket-basis, respectively. The tensor product is just $\underline{n}\otimes \underline{m}=\mathrm{spin}$-$j\otimes \mathrm{spin}$-$j'$. 

With the aid of Clebsch-Gordan coefficients (cf. \cite{pdg}), decomposition rules of tensor products into irreducible representations are straightforwardly derived,   
\be \underline{2}\otimes \underline{2} = \underline{1}\left(\fr{12-21}{\sqrt{2}}\right)\oplus \underline{3}\left(11,\fr{12+21}{\sqrt{2}},22\right),\nn\ee
\bea
\underline{3}\otimes \underline{3}&=&\underline{1}\left(\fr{13-22+31}{\sqrt{3}}\right)\oplus \underline{3}\left(\fr{12-21}{\sqrt{2}},\fr{13-31}{\sqrt{2}},\fr{23-32}{\sqrt{2}}\right)\crn
&&\oplus \underline{5}\left(11,\fr{12+21}{\sqrt{2}},\fr{13+2.22+31}{\sqrt{6}},\fr{23+32}{\sqrt{2}},33\right),\nn\eea
\be 
\underline{2}\otimes \underline{3} = \underline{2}\left(\fr{\sqrt{2}.21-12}{\sqrt{3}},\fr{22-\sqrt{2}.13}{\sqrt{3}}\right)\oplus \underline{4}\left(11,\fr{21+\sqrt{2}.12}{\sqrt{3}},\fr{\sqrt{2}.22+13}{\sqrt{3}},23\right),\nn\ee
\bea 
\underline{4}\otimes \underline{4}&=&\underline{1}\left(\fr{14-23+32-41}{2}\right)\crn
&&\oplus \underline{3}\left(\fr{\sqrt{3}.13-2.22+\sqrt{3}.31}{\sqrt{10}},\fr{3.14-23-32+3.41}{2\sqrt{5}},\fr{\sqrt{3}.24-2.33+\sqrt{3}.42}{\sqrt{10}}\right)\crn
&&\oplus \underline{5}\left(\fr{12-21}{\sqrt{2}},\fr{13-31}{\sqrt{2}},\fr{14+23-32-41}{2},\fr{24-42}{\sqrt{2}},\fr{34-43}{\sqrt{2}}\right)\crn
&&\oplus \underline{7}\left(11,\fr{12+21}{\sqrt{2}},\fr{13+\sqrt{3}.22+31}{\sqrt{5}},\fr{14+3.23+3.32+41}{2\sqrt{5}},\fr{24+\sqrt{3}.33+42}{\sqrt{5}},\fr{34+43}{\sqrt{2}},44\right),\nn\eea 
\bea
\underline{2}\otimes \underline{4} &=& \underline{3}\left(\fr{-12+\sqrt{3}.21}{2},\fr{-13+22}{\sqrt{2}},\fr{-\sqrt{3}.14+23}{2}\right)\crn
&&\oplus \underline{5}\left(11,\fr{\sqrt{3}.12+21}{2},\fr{13+22}{\sqrt{2}},\fr{14+\sqrt{3}.23}{2},24\right),\nn\eea
\bea 
\underline{3}\otimes \underline{4} &=&\underline{2}\left(\fr{\sqrt{3}.31-\sqrt{2}.22+13}{\sqrt{6}},\fr{32-\sqrt{2}.23+\sqrt{3}.14}{\sqrt{6}}\right)\crn
&&\oplus\underline{4} \left(\fr{\sqrt{3}.21-\sqrt{2}.12}{\sqrt{5}},\fr{\sqrt{6}.31+22-2\sqrt{2}.13}{\sqrt{15}},\fr{2\sqrt{2}.32-23-\sqrt{6}.14}{\sqrt{15}},\fr{\sqrt{2}.33-\sqrt{3}.24}{\sqrt{5}}\right)\crn
&&\oplus\underline{6}\left(11,\fr{\sqrt{2}.21+\sqrt{3}.12}{\sqrt{5}},\fr{31+\sqrt{6}.22+\sqrt{3}.13}{\sqrt{10}},\fr{\sqrt{3}.32+\sqrt{6}.23+14}{\sqrt{10}},\fr{\sqrt{3}.33+\sqrt{2}.24}{\sqrt{5}},34\right),\nn\eea
\bea
\underline{5}\otimes \underline{5}&=& \underline{1}\left(\fr{15-24+33-42+51}{\sqrt{5}}\right)\crn
&&\oplus\underline{3}\left(\fr{\sqrt{2}.14-\sqrt{3}.23+\sqrt{3}.32-\sqrt{2}.41}{\sqrt{10}},\fr{2.15-24+42-2.51}{\sqrt{10}},\right.\crn
&&\left.\fr{\sqrt{2}.25-\sqrt{3}.34+\sqrt{3}.43-\sqrt{2}.52}{\sqrt{10}}\right)\crn
&&\oplus \underline{5}\left(\fr{\sqrt{2}.13-\sqrt{3}.22+\sqrt{2}.31}{\sqrt{7}},\fr{\sqrt{6}.14-23-32+\sqrt{6}.41}{\sqrt{14}},\fr{2.15+24-2.33+42+2.51}{\sqrt{14}},\right.\crn
&&\left.\fr{\sqrt{6}.25-34-43+\sqrt{6}.52}{\sqrt{14}},\fr{\sqrt{2}.35-\sqrt{3}.44+\sqrt{2}.53}{\sqrt{7}}\right)\crn
&&\oplus\underline{7}\left(\fr{12-21}{\sqrt{2}},\fr{13-31}{\sqrt{2}},\fr{\sqrt{3}.14+\sqrt{2}.23-\sqrt{2}.32-\sqrt{3}.41}{\sqrt{10}},\fr{15+2.24-2.42-51}{\sqrt{10}},\right.\crn
&&\left.\fr{\sqrt{3}.25+\sqrt{2}.34-\sqrt{2}.43-\sqrt{3}.52}{\sqrt{10}},\fr{35-53}{\sqrt{2}},\fr{45-54}{\sqrt{2}}\right)\crn
&&\underline{9}\left(11,\fr{12+21}{\sqrt{2}},\fr{\sqrt{3}.13+2\sqrt{2}.22+\sqrt{3}.31}{\sqrt{14}},\fr{14+\sqrt{6}.23+\sqrt{6}.32+41}{\sqrt{14}},\right.\crn
&&\left.\fr{15+4.24+6.33+4.42+51}{\sqrt{70}},\fr{25+\sqrt{6}.34+\sqrt{6}.43+52}{\sqrt{14}},\right.\crn
&&\left.\fr{\sqrt{3}.35+2\sqrt{2}.44+\sqrt{3}.53}{\sqrt{14}},\fr{45+54}{\sqrt{2}},55\right),\nn\eea and so forth for $\underline{2}\otimes \underline{5}$, $\underline{3}\otimes \underline{5}$, and $\underline{4}\otimes \underline{5}$. To be concrete, for instance $x=(x_1,x_2)$ and $y=(y_1,y_2)$, we have $xy=(xy)_{\underline{1}}\oplus (xy)_{\underline{3}}$, where \bea (xy)_{\underline{1}} &=&(x_1y_2-x_2y_1)/\sqrt{2},\\ 
 (xy)_{\underline{3}} &=& (x_1 y_1, (x_1y_2+x_2y_1)/\sqrt{2},x_2y_2). \eea 
 
It is noted that since all $SU(2)$ representations are real, we need not necessarily consider their conjugated representations. If a conjugated representation exists, by contrast, we can transform it to the normal one, say $\underline{2}=\ep \underline{2}^*$, $\underline{3}=\ep' \underline{3}^*$, and so forth, where 
\be \ep=\begin{pmatrix} 0 & 1\\
-1 & 0\end{pmatrix},\hs \ep'=\begin{pmatrix}0 & 0 & 1\\
0 & -1 & 0\\
1 & 0 & 0\end{pmatrix}, \ee and so forth, and the above rules apply.

The overall factors on resultant irreducible representations, e.g. $1/\sqrt{2}$ in $(xy)_{\underline{1}}$, which are field normalization (exactly arising from orthonormalized bases) coefficients can be conveniently omitted or not. This work uses the full forms given above for the Yukawa Lagrangian and scalar potential, since the overall factors do not make scene as possibly absorbed into the mass and coupling parameters. However, for the kinetic term, including its gauge interaction, such overall factors are suppressed, in order to keep the canonical form.

\section{\label{appb} Lagrangian for the model with $n=m=5$}

For this model, the Lagrangian relevant to the new fields, including all scalars, takes the form, \bea \mathcal{L} &\supset& \bar{\xi} i \ga^\mu \mathcal{D}_\mu \xi+\sum_S (\mathcal{D}^{\mu} S)^\dagger (\mathcal{D}_{\mu} S) -(1/4) A'_{j\mu\nu} A'^{\mu\nu}_j\crn
&&+\left[h \bar{l}_{L} \Phi \xi-(1/2) (M+f \varphi)\xi\xi+H.c.\right]-V(S), \eea where $S=\{H,\Phi,\varphi\}$,  $\mathcal{D}_{\mu}=\pa_\mu + ig T_j A_{j\mu}+i g_B Y B_\mu+ i g' T'_j A'_{j\mu}$ is the covariant derivative as coupled to $SU(2)_L\otimes U(1)_Y\otimes SU(2)'_L$, and $A'_{j\mu\nu}=\pa_\mu A'_{j\nu}-\pa_\nu A'_{j\mu}-g'\ep_{jkl} A'_{k\mu}A'_{l\nu}$ is $SU(2)'_L$ field strength. Notice that the adjoint dark gauge boson takes the form $A'=T'_j A'_j\sim (A'^+,A'^0,A'^-)$ where the last one is given in the basis of $T'_3$ eigenstates, i.e. arranged in the $T'_3$ weight order, with $A'^\pm\equiv (A'_1\mp i A'_2)/\sqrt{2}$ and $A'^0\equiv A'_3$. Here, the superscripts $^{\pm,0}$ label only $T'_3$ values (i.e., weights), not electric charge; indeed, their electric charge is zero. 

The scalar potential is 
\bea V(S) &=& \mu^2_1\tilde{H}H+\mu^2_2\tilde{\Phi}\Phi +\mu^2_3 \tilde{\varphi}\varphi \crn
&&+[\mu_4 \tilde{\Phi}\Phi \varphi +\mu_5 H\Phi\varphi +H.c.]\crn
&&+\la_1(\tilde{H}H)^2+\la_2(\tilde{\Phi}\Phi)^2+\la_3(\tilde{\varphi}\varphi)^2\crn
&&+\la_4 (\tilde{H} H)(\tilde{\Phi} \Phi)+(\la_5\tilde{H}H+\la_6\tilde{\Phi}\Phi)(\tilde{\varphi}\varphi)\crn
&&+\la_7 (\tilde{H}\Phi)(\tilde{\Phi}H)+\la_8(\tilde{\Phi}\varphi)(\tilde{\varphi}\Phi)\crn
&&+[\la_9(H\Phi)+H.c.]\tilde{\varphi}\varphi+\la_{10}(\tilde{H}\tilde{\Phi})(H\Phi)\crn
&&+[\la_{11} (H\Phi)(H\Phi)+H.c.], \eea where $\tilde{H}=\ep H^*$ and $\tilde{\Phi}=\ep \Phi^* \ep'$, with $\ep,\ep'$ supplied in Appendix \ref{app}. The soft terms $\mu_{4,5}$ and the couplings $\la_{9,11}$ are generally complex. However, they can be considered to be real for the following computation, since otherwise their phases can be removed by redefining the relevant fields. Additionally, the combinations of types $\varphi\varphi$ and $\tilde{\varphi}\tilde{\varphi}$ are possible, in addition to the canonical form $\tilde{\varphi}\varphi$. However, they violate a phase transformation $e^{i x}$ unlike $\tilde{\varphi}\varphi$. Furthermore, we might have many/alternative possibilities for constructing an invariant tensor product, e.g. \bea \la_6(\tilde{\Phi}\Phi)(\tilde{\varphi}\varphi)&\rightarrow& \la^{(1)}_6(\tilde{\Phi}\Phi)_{\underline{1}}(\tilde{\varphi}\varphi)_{\underline{1}}+\la^{(3)}_6(\tilde{\Phi}\Phi)_{\underline{3}}(\tilde{\varphi}\varphi)_{\underline{3}}+\la^{(5)}_6(\tilde{\Phi}\Phi)_{\underline{5}}(\tilde{\varphi}\varphi)_{\underline{5}}\crn
&&+\la^{(7)}_6(\tilde{\Phi}\Phi)_{\underline{7}}(\tilde{\varphi}\varphi)_{\underline{7}}+\la^{(9)}_6(\tilde{\Phi}\Phi)_{\underline{9}}(\tilde{\varphi}\varphi)_{\underline{9}}.\eea Obviously, not all of the tensor combinations are independent and that the physics with minimal couplings by smallest dimension decompositions is the most relevant. In other words, the non-minimal couplings if viable will be not interpreted for the above potential.

To let the potential be bounded from below and achieve the relevant vacuum structure for scalar fields, the potential parameters must obey
\be \la_{1,2,3}>0,\hs \mu^2_{1,2,3}<0. \ee Here, the conditions for $\la_{1,2,3}$ are determined if $V(S)>0$ for $S=H,\Phi,\varphi$ separately tending to infinity. Additionally, the supplemental conditions for $V(S)>0$ when two of the scalar fields simultaneously tending to infinity are \bea &&\la_4+(\la_7+\la_{10}+2\la_{11})\Theta(-\la_7-\la_{10}-2\la_{11})>-2\sqrt{\la_1\la_2},\\
&&\la_6+\la_8\Theta(-\la_8)>-2\sqrt{\la_2\la_3},\hs \la_5>-2\sqrt{\la_1\la_3},
\eea where $\Theta(x)$ is the Heaviside step function. Note that the conditions for $V(S)>0$ when three of the scalar fields simultaneously tending to infinity would supply extra conditions. Additionally, the conditions for physical scalar masses squared to be positive may be also presented. All such constraints are skipped for brevity.              

At the minimum of the potential energy, we obtain the condition         
\bea
0&=& \mu_1^2+\la_1 v^2-\frac{\la_{10}}{2\sqrt{5}}\left(u_1^2+u_3^3+u_5^2\right)+\frac{\la_{11}}{\sqrt{5}}\left( u_3^2+2u_1u_5\right)\crn
&& -\frac{\la_{4}}{2\sqrt{5}}\left(u_1^2+u_3^3+u_5^2\right)-\frac{\la_5 }{2\sqrt{5}}\La^2-\frac{\la_9 u_3 }{\sqrt{70}v}\La^2 -\frac{u_5}{\sqrt{10}v} \mu_5 \La, \eea \bea  
0&=& \mu_2^2+\frac{\la_2}{\sqrt{5}}\left(u_1^2+u_3^2+u_5^2 \right)-\frac{\la_4}{2}v^2+\frac{\la_6}{2\sqrt{5}}\La^2-\frac{\la_{10}}{2}v^2\crn
&&+\la_{11}\frac{u_5}{u_1}v^2+\sqrt{\frac{5 }{7}}\frac{u_3 }{u_1}\mu_4\La,\eea \bea 
0&=& \mu_2^2+\frac{\la_2}{\sqrt{5}}\left(u_1^2+u_3^2+u_5^2 \right)-\frac{\la_4}{2}v^2+\frac{\la_6}{2\sqrt{5}}\La^2 -\frac{v}{\sqrt{14}u_3}\la_9 \La^2\crn
&&-\frac{\la_{10}}{2}v^2+\la_{11}v^2 +\sqrt{\frac{5}{7}}\frac{u_1+u_5}{u_3} \mu_4 \La, \eea \bea
0&=&\mu_2^2+\frac{\la_2}{\sqrt{5}}\left(u_1^2+u_3^2+u_5^2 \right)-\frac{\la_4}{2}v^2+\frac{\la_6}{2\sqrt{5}}\La^2 -\frac{\la_{10}}{2}v^2\crn
&&+\la_{11}\frac{u_1}{u_5}v^2+\sqrt{\frac{5}{7}}\frac{u_3}{u_5}\mu_4 \La -\frac{v}{\sqrt{2}u_5}\mu_5 \La,\eea \bea 0&=&\mu_3^2 +\frac{\la_3}{\sqrt{5}}\La^2-\frac{\la_5}{2}v^2 +\frac{\la_6}{2\sqrt{5}}\left(u_1^2+u_3^2+u_5^2\right)
-\sqrt{\frac{2}{7}} \la_9 u_3 v\crn
&&+ \sqrt{\frac{5}{7}} \frac{u_3}{\La}\left(u_1+u_5 \right)\mu_4- \frac{u_5}{\sqrt{2}\La} \mu_5 v.
\eea The five equations always give a solution of $(\La,v,u_{1,3,5})$ in which $\La$ is governed by $|\mu_3|$ scale, while $v,u_{1,3,5}$ are by $|\mu_{1,2}|$ scales, with appropriately-adjusting scalar self-couplings and $\mu_{4,5}$. Although we do not deal with the issues of $v,u_{1,3,5}\ll \La$ and $m_H\ll m_{\Phi}$ in detail, these hierarchies are typically only $1\ \mathrm{TeV}/100\ \mathrm{GeV}\sim 10$, a necessary fine-tuning between the mass parameters as well as the dimensionless couplings easily supply an expected solution.

To proceed further, we define
$\varphi_1=\left( \La+S_{1\varphi}+iA_{1\varphi}\right)/\sqrt{2}$, $\varphi_a=\left(S_{a\varphi}+iA_{a\varphi}\right) /\sqrt{2}$, for $a=2,3,4,5$, $\Phi^0_b=\left(u_b+S_b+iA_b \right)/\sqrt{2}$, for $b=1,3,5$, and $\Phi^0_c=\left(S_c+iA_c \right)/\sqrt{2}$, for $c=2,4$. Additionally, we conveniently denote $H_1\equiv S_{1\varphi}$ for using throughout the text.  

The dark scalars $(S_2, S_4)$ mix via a mass-squared matrix as follows
\bea
-\fr 1 2 \begin{pmatrix} S_2& S_4 \end{pmatrix} \begin{pmatrix}
	m_{S_2 S_2}^2 & m_{S_2 S_4}^2 \\
	m_{S_4 S_2}^2 & m_{S_{4}S_4}^2\end{pmatrix} \begin{pmatrix} S_2 \\S_4 \end{pmatrix},
\eea
where
\bea
m_{S_{2}S_2}^2=m^2_{S_{4}S_4} &=& \frac{u^2_3-u_1u_5-u_5^2}{\left(u_1^2+u_1 u_5-u_3^2 \right)} \frac{\la_{11}v^2}{2\sqrt{5}}-\frac{1}{2\sqrt{70}}\frac{\la_9 u_3v\La^2}{\left(u_1^2+u_1 u_5-u_3^2 \right)},\\ 
m_{S_2 S_4}^2=m^2_{S_4 S_2}&=&-\frac{\la_{11}}{2\sqrt{5}}v^2-\sqrt{\frac{3}{40}}\frac{\la_{11}\left(u_1+u_5 \right)u_3v^2}{u_1^2-u_3^2+u_1u_5}+\sqrt{\frac{3}{35}}\frac{\la_9 u_1v\La^2}{4\left( u_1^2-u_3^2+u_1u_5\right)}.
\eea
We obtain the physical eigenstates 
\be S_{24}=(S_2-S_{4})/\sqrt{2},\hs S'_{24}=(S_2+S_{4})/\sqrt{2},\ee with respect to the mass eigenvalues, 
\be
 m_{S_{24}}^2=m_{S_2 S_2}^2-m_{S_2 S_4}^2, \hs m_{S'_{24}}^2 =m_{S_{2}S_2}^2+m_{S_{2}S_{4}}^2.
\ee It is clear that the scalar masses are completely separated and proportional to $\La$ scale. The lightest dark scalar is $S_{24}$ whose mass is below or above a TeV, depending on the magnitude of $\la_{9}$. 

The relevant couplings among the lightest dark scalar, the usual Higgs boson, and the new Higgs fields  are computed, collected in Table \ref{tab3}.
\begin{table}[h]
\bc
\begin{tabular}{lc}
\hline \hline
Vertex & Coupling \\
\hline 
$S_{24} S_{24} H H$ & $\frac{1}{16\sqrt{5}}\left(2 \la_{11}-\la_4-\la_{10}  \right)$ \\
$S_{24} S_{24} H$ & $\frac{1}{4\sqrt{5}}\left(2 \la_{11}-\la_4-\la_{10} \right) v $\\ 
$S_{24} S_{24} S_1$ & $\frac{\la_2}{10}u_1 $\\ 
$S_{24} S_{24} S_3$ & $\frac{\la_2}{10}u_3$ \\ 
$S_{24} S_{24} S_5$ &$ \frac{\la_2}{10}u_5$ \\  
$S_{24} S_{24} H_{1}$ & $\frac{\la_6}{20}\La+ \frac{\sqrt{3}\left(u_1-u_5\right)}{4\sqrt{10}\left( u_1^2+u_1u_5-u_3^2\right)\La}\la_{11}u_3v^2-\frac{\sqrt{3}}{8\sqrt{35}\left(u_1^2+u_1u_5-u_3^2 \right)} \la_9 u_1v \La $\\ 
$S_{24} S_{24} S_{3\varphi}$ & $\frac{u_1-u_5}{4\sqrt{20}\left( u_1^2+u_1u_5-u_3^2\right)\La}\la_{11}u_3v^2-\frac{1}{8\sqrt{70}\left(u_1^2+u_1u_5-u_3^2 \right)} \la_9 u_1v \La $ \\ 
$S_{24} S_{24} S_{5\varphi}$ & $\frac{\sqrt{3}\left(u_1-u_5\right)}{8 \sqrt{10}\left(u_1^2+u_1u_5-u_3^2 \right)\La} \la_{11}u_3 v^2 -\frac{\sqrt{3}}{8\sqrt{140}\left(u_1^2+u_1u_5-u_3^2\right)}\la_9 u_1v \La$\\ 
$H H H$ & $\frac{1}{3}\la_1v$ \\ 
$HH H_1 $ & $-\frac{1}{4\sqrt{5}} \la_5 \La $\\ 
$HHS_{3\varphi} $ & $0$ \\ 
$HHS_{5\varphi} $ & $0$ \\ 
$HHS_1 $ & $\frac{1}{4 \sqrt{5}}\left\{2 \la_{11}u_5 -\left(\la_4+\la_{10}\right)u_1 \right\}$ \\ 
$HHS_3 $ & $\frac{1}{4 \sqrt{5}}\left(2 \la_{11}-\la_{10}-\la_4 \right)u_3$ \\ 
$HHS_5 $ & $\frac{1}{4 \sqrt{5}}\left\{2 \la_{11}u_1-\left(\la_{10}+\la_4 \right)u_1\right\}$ \\ 
\hline\hline		
\end{tabular}
\caption[]{\label{tab3} Couplings of $S_{24}$ with various Higgs fields as well as those of the usual and new Higgs fields, where note that the couplings of $S_{24}$ with CP-odd scalar components vanish.}
\ec	
\end{table}

\end{document}